\documentclass[aps,prapplied,twocolumn,longbibliography]{revtex4-2}
\usepackage{graphicx}
\usepackage{dcolumn}
\usepackage{hyperref}
\usepackage{bm}
\usepackage{hyperref}
\usepackage{lipsum}
\usepackage{subcaption}  
\usepackage{caption}         
\usepackage{tikz}
\usepackage{ragged2e}
\usetikzlibrary{positioning, arrows.meta}  
\usepackage{relsize}  
\usepackage{overpic}  
\usepackage{placeins}
\usepackage{amssymb}
\usepackage{amsmath}
\usepackage{algorithm}
\usepackage{algpseudocode}
\usepackage[english]{babel}
\usepackage{makecell}
\usepackage{bm}
\usepackage{placeins}
\usepackage[export]{adjustbox}
\usepackage{xcolor}
\usepackage{caption}
\begin{document}

\title{Astroid Spinodal Boundary in Phase-Based Ising Machines}

\author{Malihe Farasat\textsuperscript{1}, Nikhil Shukla\textsuperscript{1}}
\affiliation{$^1${Department of Electrical and Computer Engineering}, University of Virginia, Charlottesville, VA {22903}, USA
}%


\begin{abstract}

Oscillator Ising machines (OIMs) and dynamical Ising machines (DIMs) encode binary spins in phase states stabilized by second-harmonic injection (SHI). In a coupled network, the competition between SHI and the instantaneous local network field reshapes each oscillator's conditional energy landscape. We show that this competition drives a transition between monostable and bistable regimes through an astroid spinodal boundary. Near this boundary, the barrier scales as $\Delta E_i\propto\mu_i^{3/2}$ at generic smooth points and as $\Delta E_i\propto\mu_i^{2}$ at the longitudinal cusp. OIMs and DIMs obey the same spinodal geometry, with their conditional landscapes related by a reversal of the transverse field. Finally, the first-harmonic conditional landscape is mathematically equivalent, up to an additive constant, to the Stoner--Wohlfarth energy of a uniaxial magnetic particle.
\end{abstract}

\maketitle

\section{Introduction}

The increasing difficulty of scaling conventional digital computing architectures, together with the growing centrality of large-scale combinatorial optimization problems (COPs) in artificial intelligence, machine learning, and data analytics, has intensified the search for alternate computational paradigms capable of accelerating such problems. One promising approach is the design of physical computing systems whose intrinsic dynamics directly implement the minimization of objective functions associated with COPs. In this context, Ising machines—hardware platforms inspired by the classic Ising model of interacting spins have emerged as a compelling class of non-von Neumann accelerators that exploit collective physical dynamics to search for low-energy configurations of the Ising Hamiltonian~\cite{mohseni2022ising,zhang2024review}. The Ising Hamiltonian, is given by,
\begin{equation}
H(\mathbf{s}) = -\frac{1}{2}\sum_{\substack{i,j\\i\neq j}} W_{ij} s_i s_j, \quad s_i \in \{+1, -1\}, \label{eq:IsingH}
\end{equation}
where, $s_i$ denotes the binary spin state of node $i$, and $W_{ij}$ defines the interaction between spins $i$ and $j$. Minimizing the Ising Hamiltonian optimally is a well-known NP-hard problem. Moreover, since many other NP-hard problems can be mapped to the minimization of $H$; Ising machines, in principle, can provide a powerful hardware framework for accelerating the broader class of COPs~\cite{lucas2014ising}.

A wide range of Ising machine implementations have been explored across quantum~\cite{chang2024quantum}, optical~\cite{mcmahon2016fully}, mechanical~\cite{ezawa2022ising}, spintronic~\cite{grimaldi2023evaluating}, and electronic platforms~\cite{chou2019analog,mallick2023cmos,bashar2021experimental, moy20221, afoakwa2021brim, Cassella}. Within this broad landscape, oscillator-based analog Ising machines constitute an important class. In this work, we focus on two phase-based implementations: Kuramoto-oscillator-based Ising machines~\cite{ wang2019oim}, which we refer to as oscillator Ising machines (OIMs), and the recently proposed dynamical Ising machines (DIMs)~\cite{ekanayake2025different}. Both architectures use second-harmonic injection (SHI) to favor the phase states $\theta_i\in\{0,\pi\}$, thereby enabling the binary spin readout $s_i=\cos\theta_i\in\{+1,-1\}$. They differ, however, in the form of their network interaction: OIMs employ phase-difference coupling $(\theta_i-\theta_j)$, whereas DIMs employ phase-sum coupling $(\theta_i+\theta_j)$.

SHI is intended to endow each oscillator with two spin-encoding phase states. In a coupled network, however, whether both states are locally stable is determined jointly by SHI and the instantaneous field generated by the neighboring oscillators. Thus, the competition between SHI and network feedback determines whether the conditional landscape is monostable or bistable. In the bistable regime, one of the wells may correspond to the spin state disfavored by the local Ising interaction.

Here, we develop an energy-landscape description of bistability in phase-based Ising machines. We show that the competition between SHI and the instantaneous network field determines whether the local conditional landscape of an
oscillator is monostable or bistable. From this landscape, we derive the spinodal boundary for the emergence of a metastable well, determine the corresponding SHI threshold, and characterize the growth of the resulting energy barrier. Moreover, we note that the conditional landscape has the same functional form as the Stoner--Wohlfarth energy of a uniaxial magnetic particle. We further show that OIMs and DIMs share the same spinodal geometry and extend the analysis to phase-encoded gradient flows containing higher odd-harmonic network interactions.
\begin{figure*}
    \centering  
    \includegraphics[width=1\linewidth]{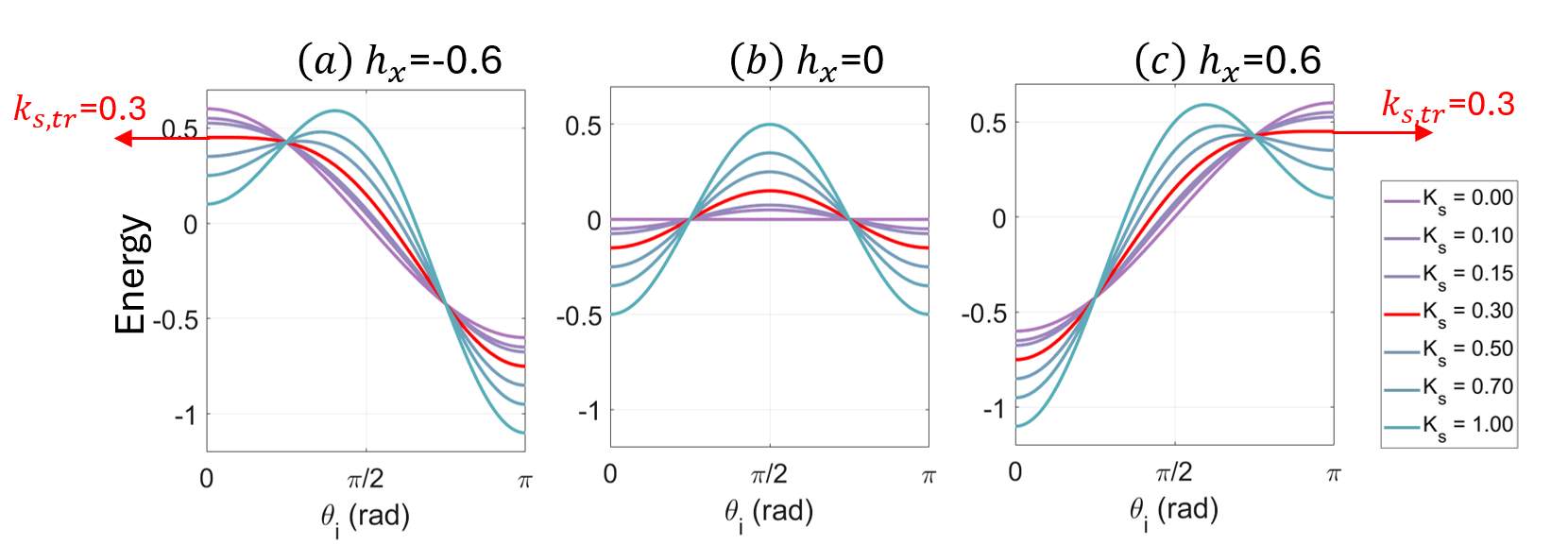}
       
        \caption{\justifying Energy profiles of the longitudinal-axis conditional landscape $E_i(\theta_i)=-h_{x,i}\cos\theta_i-(K_{\mathrm{s}}/2)\cos(2\theta_i)$
    for representative values of the local Ising field $h_{x,i}$ of (a) -0.6, (b) 0 and (c) 0.6, and SHI strength $K_{\mathrm{s}}$. For $h_{x,i}\neq 0$, increasing $K_{\mathrm{s}}$ past     $K_{\mathrm{s},\mathrm{tr},i}=|h_{x,i}|/2$ creates a metastable well at the Ising-disfavored binary state. For $h_{x,i}=0$, the two binary states are degenerate and SHI ($K_{\mathrm{s}}>0$) produces a symmetric double-well landscape.}
    \label{fig:energybarrier}
\end{figure*}

\section{Local Energy Landscape and Barrier-Mediated Pinning}
\label{sec:spinodal_freezing}

We begin by considering the instantaneous conditional energy of oscillator $i$, which combines the network interaction with the local SHI contribution. For OIMs and DIMs, this energy can be written in the unified form

\begin{equation}
E_i^{(\sigma)}(\theta_i)
=
-K\sum_{\substack{j=1\\j\neq i}}^N
W_{ij}\cos(\theta_i+\sigma\theta_j)
-\frac{K_{\mathrm{s}}}{2}\cos(2\theta_i),
\label{eq:unified_local_energy}
\end{equation}
where $\sigma=-1$ corresponds to the conventional OIM phase-difference
interaction, and $\sigma=+1$ corresponds to the DIM phase-sum interaction.

Expanding the interaction term yields
\begin{equation}
E_i^{(\sigma)}(\theta_i)
=
-h_{x,i}\cos\theta_i
-h_{y,i}^{(\sigma)}\sin\theta_i
-\frac{K_{\mathrm{s}}}{2}\cos(2\theta_i),
\label{eq:local_field_energy}
\end{equation}
with
\begin{equation}
h_{x,i}
=
K\sum_{\substack{j=1\\j\neq i}}^N W_{ij}\cos\theta_j,
\qquad
h_{y,i}^{(\sigma)}
=
-\sigma K\sum_{\substack{j=1\\j\neq i}}^N W_{ij}\sin\theta_j .
\label{eq:hx_hy_definitions}
\end{equation}
The quantities $h_{x,i}$ and $h_{y,i}^{(\sigma)}$ are therefore the
longitudinal and transverse components of the instantaneous local network
field acting on oscillator $i$. The
longitudinal direction coincides with the Ising axis defined by the binary
phases $\theta_i\in\{0,\pi\}$. The longitudinal component $h_{x,i}$ is the Ising-relevant bias because it sets the energy difference between the two binary phase states:
\begin{equation}
E_i^{(\sigma)}(0)-E_i^{(\sigma)}(\pi)=-2h_{x,i}.
\label{eq:longitudinal_bias}
\end{equation}

\noindent Thus, the sign of $h_{x,i}$ selects which of the two binary states is lower in energy, while its magnitude sets the corresponding energy splitting.

The transverse component $h_{y,i}^{(\sigma)}$ does not contribute to this binary-state energy splitting, since $\sin 0=\sin\pi=0$. It nevertheless enters the continuous conditional landscape and affects the local phase dynamics. In particular, 
\begin{equation}
\left.
\frac{\partial E_i^{(\sigma)}}{\partial \theta_i}
\right|_{\theta_i=0}
=
-h_{y,i}^{(\sigma)},
\qquad
\left.
\frac{\partial E_i^{(\sigma)}}{\partial \theta_i}
\right|_{\theta_i=\pi}
=
h_{y,i}^{(\sigma)} .
\label{eq:transverse_binary_force}
\end{equation}
Thus, when $h_{y,i}^{(\sigma)}\neq 0$, the binary phases are generally not stationary points of the conditional landscape. Instead, the transverse field displaces the extrema away from the binary axis, modifies the barrier structure, and changes the spinodal conditions for the creation and destruction of local energy wells. This separation between the Ising bias set by $h_{x,i}$ and the off-axis landscape deformation induced by $h_{y,i}^{(\sigma)}$ is central to the energetic interpretation developed below.

The same decomposition also relates the conditional energy landscapes of OIMs
and DIMs. For a fixed snapshot of the neighboring phases, the two systems have
identical longitudinal fields but opposite transverse fields:
\begin{equation}
h_{x,i}^{\mathrm{DIM}}=h_{x,i}^{\mathrm{OIM}},
\qquad
h_{y,i}^{\mathrm{DIM}}=-h_{y,i}^{\mathrm{OIM}}.
\label{eq:OIM_DIM_mirror}
\end{equation}
Thus, the corresponding conditional landscapes are related by reflection in the transverse-field direction. Their actual transient trajectories, however, need not coincide, because the neighboring phases evolve under different phase-difference and phase-sum coupling laws.

The stationary points of the conditional landscape in
Eq.~\eqref{eq:local_field_energy} satisfy
\begin{equation}
\frac{\partial E_i^{(\sigma)}}{\partial \theta_i}
=
h_{x,i}\sin\theta_i
-h_{y,i}^{(\sigma)}\cos\theta_i
+K_{\mathrm{s}}\sin(2\theta_i)
=0,
\label{eq:first_derivative_field}
\end{equation}
and the corresponding local curvature is
\begin{equation}
{E_i''(\theta_i)}^{(\sigma)}
=
h_{x,i}\cos\theta_i
+h_{y,i}^{(\sigma)}\sin\theta_i
+2K_{\mathrm{s}}\cos(2\theta_i).
\label{eq:second_derivative_field}
\end{equation}
A local well is created or destroyed at a spinodal point, where a local minimum
coalesces with the adjacent maximum and the curvature of the conditional
landscape vanishes. The spinodal boundary is therefore obtained by imposing
Eqs.~\eqref{eq:first_derivative_field} and
\eqref{eq:second_derivative_field} with
$\partial^2 E_i^{(\sigma)}/\partial\theta_i^2=0$.

\subsection{Purely longitudinal field: onset of local bistability}
\label{sec:axis_case}
We first consider the special case of a purely longitudinal local network
field, for which $h_{y,i}^{(\sigma)}=0$. A natural realization of this limit occurs when the neighboring oscillators have already binarized, $\theta_j\in\{0,\pi\}$, so that $\sin\theta_j=0$ and $s_j=\cos\theta_j=\pm1$. The transverse field then vanishes, and the longitudinal component reduces to the usual local Ising bias,
\begin{equation}
h_{x,i}
=
K\sum_{\substack{j=1\\j\neq i}}^N W_{ij}s_j .
\label{eq:local_ising_field}
\end{equation}
In this limit, the conditional landscape seen by oscillator $i$ is governed
only by the Ising-relevant local bias and the SHI-induced bistabilizing term:
\begin{equation}
E_i(\theta_i)
=
-h_{x,i}\cos\theta_i
-\frac{K_{\mathrm{s}}}{2}\cos(2\theta_i).
\label{eq:axis_energy}
\end{equation}
The corresponding stationary condition follows from
Eq.~\eqref{eq:first_derivative_field} and factorizes as
\begin{equation}
\frac{\partial E_i}{\partial \theta_i}
=
h_{x,i}\sin\theta_i
+
K_{\mathrm{s}}\sin(2\theta_i)
=
\sin\theta_i
\left(
h_{x,i}+2K_{\mathrm{s}}\cos\theta_i
\right).
\label{eq:axis_force_factored}
\end{equation}

The factor $\sin\theta_i=0$ gives the two binary stationary points $\theta_i\in\{0,\pi\}$, which are present for all values of $K_{\mathrm{s}}$ and $h_{x,i}$. The second factor gives an additional pair of stationary points satisfying 
\begin{equation}
\cos\theta_i=-\frac{h_{x,i}}{2K_{\mathrm{s}}},
\label{eq:extra_root}
\end{equation}
which exist only when
\begin{equation}
2K_{\mathrm{s}}>|h_{x,i}|.
\label{eq:axis_threshold}
\end{equation}

Thus, Eq.~\eqref{eq:axis_threshold} marks the onset of local bistability in the on-axis landscape.

The nature of this transition follows from the curvature,
\begin{equation}
E_i''(\theta_i)
=
h_{x,i}\cos\theta_i
+
2K_{\mathrm{s}}\cos(2\theta_i).
\end{equation}
Taking $h_{x,i}>0$ without loss of generality, the state $\theta_i=0$ ($s_i=+1$) is favored by the local Ising bias. Its curvature is $E_i''(0)=h_{x,i}+2K_{\mathrm{s}}>0$, so it remains a local minimum for all $K_{\mathrm{s}}>0$. The antipodal binary state has curvature
\begin{equation}
E_i''(\pi)=2K_{\mathrm{s}}-h_{x,i}.
\end{equation}
Therefore, for $2K_{\mathrm{s}}<|h_{x,i}|$, the point $\theta_i=\pi$ is a
maximum and the local landscape contains only one stable well. When $2K_{\mathrm{s}}>|h_{x,i}|$, this point becomes a local minimum. At the same threshold, the additional stationary points in
Eq.~\eqref{eq:extra_root} emerge from the binary state disfavored by the local Ising field and form the barriers separating the newly created metastable well from the Ising-favored binary state. Equivalently, as the SHI strength crosses
\begin{equation}
K_{\mathrm{s},\mathrm{tr},i}
=
\frac{|h_{x,i}|}{2}
=
\frac{K}{2}
\Big|
\sum_{\substack{j=1\\j\neq i}}^{N} W_{ij}s_j
\Big|,
\label{eq:axis_Kstr}
\end{equation}
the conditional landscape changes from a single-well potential to a double-well potential.

In the purely longitudinal limit, the threshold in Eq.~\eqref{eq:axis_Kstr} marks the point at which the binary state disfavored by the local Ising field changes from unstable to metastable. Below this threshold, the conditional landscape contains a single stable well, and the oscillator relaxes toward the Ising-favored binary state. Above the threshold, a competing well is stabilized at the ``Ising-disfavored'' binary state . For example, when $h_{x,i}>0$, Eq.~\eqref{eq:longitudinal_bias} yields $E_i(0)-E_i(\pi)=-2h_{x,i}<0$, so the binary state $s_i=+1$ is favored by the local Ising field. If the oscillator is already near $\theta_i=\pi$ when this metastable well is formed, it can remain trapped at $s_i=-1$, thereby suppressing a transition to the lower-Ising-energy state.

From a functional standpoint, bistability becomes detrimental when the oscillator is trapped in the well corresponding to the Ising-disfavored binary state. Increasing $K_{\mathrm{s}}$ deepens this local trap, whereas increasing $|h_{x,i}|$ strengthens the tilt toward the Ising-favored state and eventually restores a single-well landscape. This on-axis limit captures, at the single-oscillator level, the competition between network-driven selection of the lower-energy spin state and SHI-induced local stabilization.

Fig.~\ref{fig:energybarrier} shows the evolution of the purely longitudinal conditional landscape as the SHI strength is increased. As an illustrative example, we take $h_{x,i}=0.6$, for which the longitudinal field favors
$\theta_i=0$, while $\theta_i=\pi$ is initially unstable. At the critical value $K_{\mathrm{s,tr},i}=|h_{x,i}|/2=0.3$, the curvature at $\theta_i=\pi$ vanishes. For $K_{\mathrm{s}}>K_{\mathrm{s,tr},i}$, this state becomes a local minimum, and an intervening maximum separates it from the lower-energy minimum at $\theta_i=0$. The landscape is then bistable, with the higher-energy well at $\theta_i=\pi$ constituting the metastable state. The case $h_{x,i}=-0.6$, included as the corresponding illustrative example of opposite longitudinal bias, is related by the
transformation $\theta_i\mapsto\pi-\theta_i$, which interchanges the favored and metastable binary states. When $h_{x,i}=0$, the longitudinal bias vanishes, the two binary states are degenerate, and any $K_{\mathrm{s}}>0$ produces a symmetric double-well landscape. Thus, increasing $K_{\mathrm{s}}$ drives the transition from monostability to bistability and raises the barrier between the two wells, whereas increasing $|h_{x,i}|$ strengthens the energetic asymmetry between them.
\begin{figure*}
    \centering

     \includegraphics[width=1\linewidth]{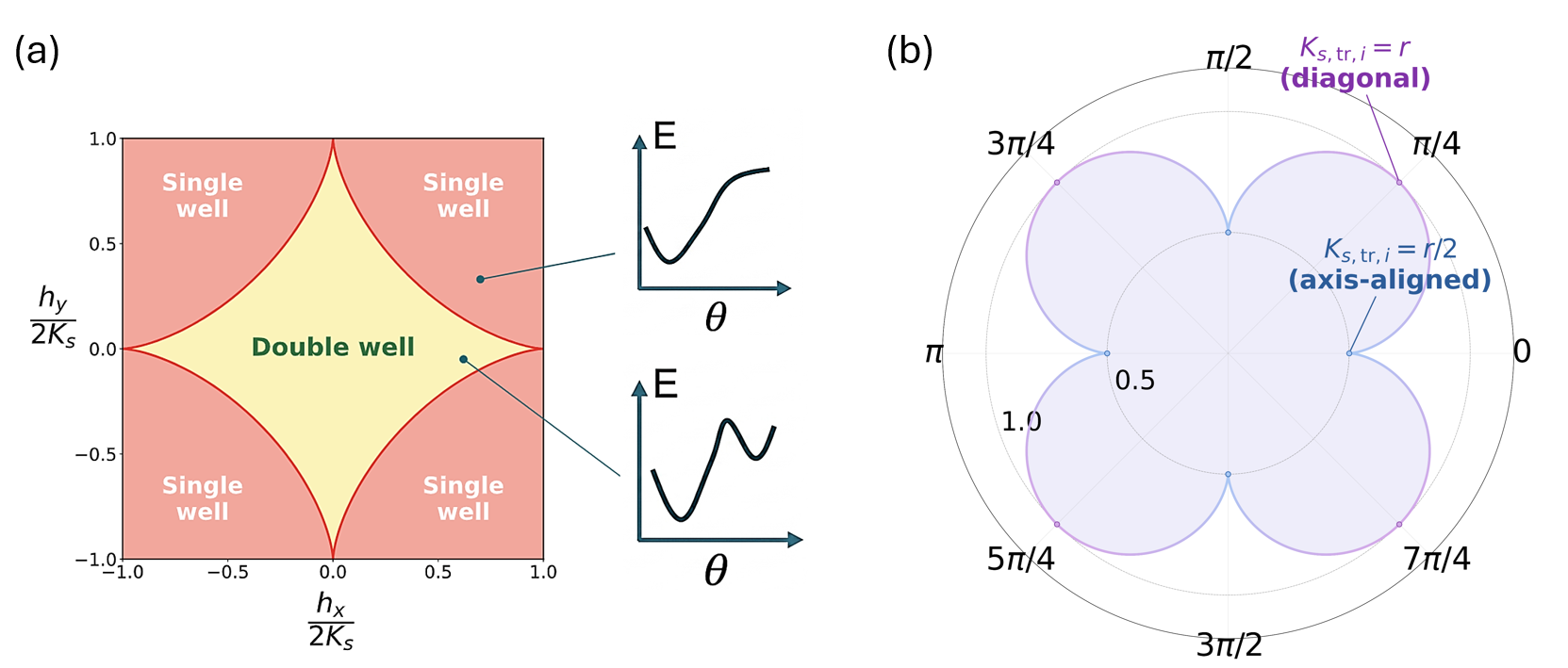}
      \centering
                  \caption{\justifying (a) Schematic of the spinodal astroid in the normalized local-field plane $\bigl(h_x/2K_{\mathrm{s}},\,h_y/2K_{\mathrm{s}}\bigr)$. The astroid boundary separates bistable conditional landscapes inside the astroid from monostable landscapes outside it. (b) Polar plot of the switching threshold $K_{s,\mathrm{tr},i}$ vs.\ field angle, showing that diagonal fields require the full astroid radius to switch ($K_{s,\mathrm{tr},i} = r$), while axis-aligned fields switch at half that threshold ($K_{s,\mathrm{tr},i} = r/2$).}
    \label{fig:Astroid}
\end{figure*}

\subsection{General field orientation: the astroid spinodal boundary}
We next consider the general case in which the local network field has a nonzero transverse component, $h_{y,i}^{(\sigma)}\neq 0$. In this setting, a local well is created or destroyed through a spinodal event, at which a local minimum coalesces with the adjacent barrier top. Since the conditional landscape is one-dimensional, this barrier top is a local maximum. The spinodal boundary is therefore obtained by imposing stationarity and vanishing curvature,
\begin{equation}
\frac{\partial E_i^{(\sigma)}}{\partial \theta_i}=0,
\qquad
\frac{\partial^2 E_i^{(\sigma)}}{\partial \theta_i^2}=0.
\label{eq:spinodal_conditions}
\end{equation}

Solving Eqs.~\eqref{eq:first_derivative_field} and
\eqref{eq:second_derivative_field} with zero curvature gives a parametric
description of the spinodal in terms of the coalescence angle
$\theta_{\mathrm{sp}}$:
\begin{equation}
h_{x,i}
=
-2K_{\mathrm{s}}\cos^3\theta_{\mathrm{sp}},
\qquad
h_{y,i}^{(\sigma)}
=
2K_{\mathrm{s}}\sin^3\theta_{\mathrm{sp}}.
\label{eq:spinodal_parametric}
\end{equation}
Eliminating $\theta_{\mathrm{sp}}$ yields the astroid-shaped boundary (Fig. \ref{fig:Astroid})
\begin{equation}
\left|\frac{h_{x,i}}{2K_{\mathrm{s}}}\right|^{2/3}
+
\left|\frac{h_{y,i}^{(\sigma)}}{2K_{\mathrm{s}}}\right|^{2/3}
=
1 .
\label{eq:spinodal_astroid}
\end{equation}
For fixed $K_{\mathrm{s}}$, field configurations satisfying the strict
inequality
\begin{equation}
\left|\frac{h_{x,i}}{2K_{\mathrm{s}}}\right|^{2/3}
+
\left|\frac{h_{y,i}^{(\sigma)}}{2K_{\mathrm{s}}}\right|^{2/3}
< 1
\end{equation}
correspond to bistable conditional landscapes, whereas configurations outside
the astroid correspond to a monostable landscape. Thus, the astroid identifies
the instantaneous field values for which the local landscape can support two
stable wells.

\begin{figure}
    \centering        
    \includegraphics[width=0.8\linewidth]{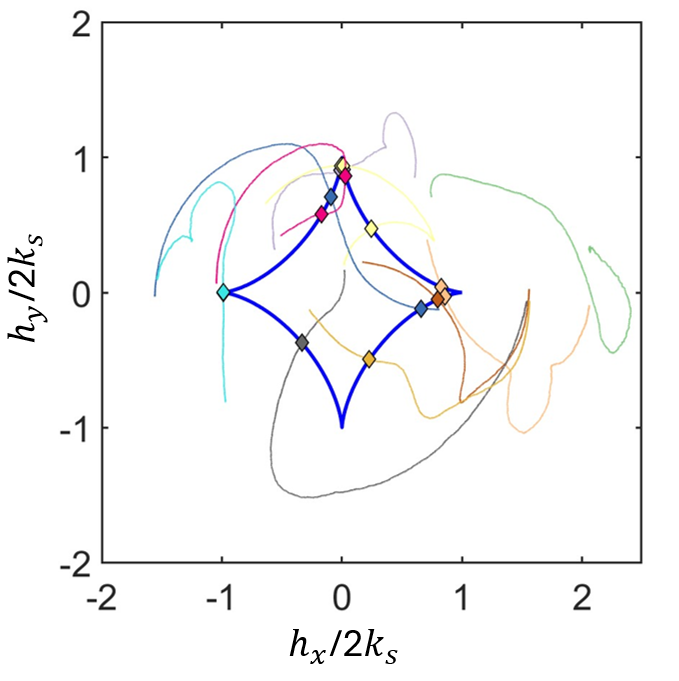}

        \caption{\justifying Trajectories of the normalized local fields $\bigl(h_{y,i}/2K_{\mathrm{s}},\,h_{x,i}/2K_{\mathrm{s}}\bigr)$ for all
oscillators in a representative randomly generated graph with $N$ nodes and edge density of p, shown relative to the spinodal astroid. Diamonds indicate crossings of the astroid boundary, at which the corresponding conditional landscape changes between monostable and bistable regimes. Simulation parameters : $K=1, K_s=1 , N=10, p=0.5$.}

    \label{fig:oimdimtrajectory}
\end{figure}

Fig.~\ref{fig:oimdimtrajectory} shows field-space trajectories for an OIM network, plotted against the spinodal astroid of Eq.~\eqref{eq:spinodal_astroid}. As the figure shows, individual nodes repeatedly cross the astroid boundary over the course of the computation, transitioning back and forth between the monostable and bistable regions. This result is generated using a random Erdős–Rényi graph with 10 nodes and edge density p = 0.5, used to realize an OIM network.

Equivalently, for a fixed local field
$(h_{x,i},h_{y,i}^{(\sigma)})$, the SHI strength required to enter the
bistable regime is
\begin{equation}
K_{\mathrm{s},\mathrm{tr},i}
=
\frac{1}{2}
\left(
|h_{x,i}|^{2/3}
+
|h_{y,i}^{(\sigma)}|^{2/3}
\right)^{3/2}.
\label{eq:Ks_transition_astroid}
\end{equation}
Eq.~\eqref{eq:Ks_transition_astroid} generalizes the on-axis threshold in Eq.~\eqref{eq:axis_Kstr}.

Although the transverse component $h_{y,i}^{(\sigma)}$ does not alter the
energy splitting between the binary states $\theta_i=0$ and $\theta_i=\pi$,
it enters the spinodal condition and therefore affects the onset of local
bistability. To isolate this orientational dependence, we parameterize the
local network field as
\[
h_{x,i}=r\cos\phi,
\qquad
h_{y,i}^{(\sigma)}=r\sin\phi,
\]
where,
\[
r=\sqrt{h_{x,i}^{2}+\left(h_{y,i}^{(\sigma)}\right)^{2}}
\]
is the field magnitude and $\phi$ is its orientation relative to the longitudinal
axis. Eq.~\eqref{eq:Ks_transition_astroid} then gives
\begin{equation}
K_{s,\mathrm{tr},i}
=
\frac{r}{2}
\left(
|\cos\phi|^{2/3}
+
|\sin\phi|^{2/3}
\right)^{3/2}.
\label{eq:threshold-orientation}
\end{equation}

For fixed $r$, the threshold is minimized for a longitudinal axis-aligned field,
\begin{equation}
\phi=n\frac{\pi}{2},
\qquad n\in\mathbb{Z},
\end{equation}
for which
\begin{equation}
K_{s,\mathrm{tr},i}
=
\frac{r}{2}.
\label{eq:threshold-min}
\end{equation}
The threshold is maximized along the diagonal directions,
\begin{equation}
\phi=(2n+1)\frac{\pi}{4},
\qquad n\in\mathbb{Z},
\end{equation}
for which
\begin{equation}
K_{s,\mathrm{tr},i}=r.
\end{equation}

Thus, fields of equal magnitude can require SHI strengths differing by a
factor of two to produce a bistable conditional landscape. This directional
dependence follows directly from the astroid geometry: its radial extent is
largest along the longitudinal and transverse axes and smallest along the
diagonals. Consequently, an axis-aligned field remains within the bistable
region at a smaller SHI strength than a diagonal field of the same magnitude.

\textit{Impact of graph structure.}
The dependence of the local bistability condition on graph structure enters
through the local phasor
\begin{equation}
h_{x,i}+i h_{y,i}^{(\sigma)}
=
K\sum_{j\neq i} W_{ij}\,e^{-i\sigma\theta_j}.
\label{eq:local_phasor}
\end{equation}
Each neighbor therefore contributes a weighted phasor to the instantaneous
network field acting on oscillator $i$. During early transients, before the
neighboring phases have developed a correlated or nearly binary pattern, these
phasors are generally misaligned. The local field is then governed not only by
the number of neighbors, but also by the degree of constructive or destructive
interference among their phasor contributions. If the neighboring phases are
treated as approximately uncorrelated, the typical field magnitude obeys
\begin{equation}
\left\langle
\left|h_{x,i}+i h_{y,i}^{(\sigma)}\right|^2
\right\rangle^{1/2}
\sim
K\left(\sum_{j\neq i} W_{ij}^2\right)^{1/2}.
\label{eq:phasor_rms}
\end{equation}
For an unweighted graph this scales as $K\sqrt{d_i}$, in which $d_i$ is  the degree of node i, that is smaller than
the coherent scale $K d_i$ obtained when all neighbor contributions are
aligned~\cite{restrepo2005onset}. Thus, the tendency of a node to enter the bistable region is not
controlled by degree alone; it also depends on the instantaneous phase
organization of its neighborhood.

\subsection{Spinodal margin and barrier growth}
To quantify the position of the instantaneous local field relative to the
spinodal boundary, we define the spinodal margin
\begin{equation}
\mu_i(t)=1-\ell_i(t),
\label{eq:spinodal_margin}
\end{equation}
where
\begin{equation}
\ell_i(t)
=
\left|\frac{h_{x,i}(t)}{2K_{\mathrm{s}}(t)}\right|^{2/3}
+
\left|\frac{h_{y,i}^{(\sigma)}(t)}{2K_{\mathrm{s}}(t)}\right|^{2/3}.
\label{eq:astroid_coordinate}
\end{equation}
Accordingly, $\mu_i>0$, $\mu_i=0$, and $\mu_i<0$ correspond to field
points inside, on, and outside the astroid, respectively. Within the
bistable region, increasing $\mu_i$ indicates motion farther from the
spinodal boundary in the normalized astroid coordinate. Near a given
spinodal crossing, $\mu_i$ therefore provides a natural control parameter
for the emergence and growth of the local energy barrier.

We next relate the spinodal margin to the barrier separating the newly formed
minimum from the adjacent local maximum. Consider first a nonaxial point of the
astroid, where the boundary is smooth and $\mu_i$ provides a regular local
coordinate transverse to it. Let
\[
q=\theta_i-\theta_{\mathrm{sp}}
\]
denote the displacement from the coalescence point. At the spinodal,
\[
\left.
\frac{\partial E_i^{(\sigma)}}{\partial\theta_i}
\right|_{\mathrm{sp}}
=
\left.
\frac{\partial^2 E_i^{(\sigma)}}{\partial\theta_i^2}
\right|_{\mathrm{sp}}
=0,
\]
whereas
$\partial_{\theta_i}^3E_i^{(\sigma)}|_{\mathrm{sp}}\neq0$
at a generic smooth point. A joint expansion in $q$ and $\mu_i$ therefore gives
\begin{equation}
\frac{\partial E_i^{(\sigma)}}{\partial\theta_i}
=
A_i\mu_i+B_iq^2
+
O\!\left(\mu_i q,q^3,\mu_i^2\right),
\label{eq:fold_force}
\end{equation}
where
\[
A_i=
\left.
\frac{\partial}{\partial\mu_i}
\frac{\partial E_i^{(\sigma)}}{\partial\theta_i}
\right|_{\mathrm{sp}},
\qquad
B_i=
\frac{1}{2}
\left.
\frac{\partial^3E_i^{(\sigma)}}{\partial\theta_i^3}
\right|_{\mathrm{sp}} .
\]
The signs of $A_i$ and $B_i$ are opposite on the bistable side of the
boundary. The minimum--maximum pair is consequently located at
\begin{equation}
q_{\pm}
=
\pm
\left(-\frac{A_i}{B_i}\right)^{1/2}
\mu_i^{1/2}
+
O(\mu_i).
\label{eq:fold_stationary_points}
\end{equation}
Integrating Eq.~\eqref{eq:fold_force} with respect to $q$ gives
\begin{equation}
E_i^{(\sigma)}(q)
=
E_{\mathrm{sp}}
+
A_i\mu_i q
+
\frac{B_i}{3}q^3
+
O\!\left(\mu_i q^2,q^4,\mu_i^2q\right).
\label{eq:fold_energy}
\end{equation}
The resulting barrier height is
\begin{equation}
\Delta E_i
=
\frac{4}{3}
\frac{|A_i|^{3/2}}{|B_i|^{1/2}}
\mu_i^{3/2}
+
o(\mu_i^{3/2}).
\label{eq:barrier_generic}
\end{equation}
Since the conditional energy is proportional to $K_{\mathrm{s}}$ when expressed
in terms of the normalized fields
$h_{x,i}/(2K_{\mathrm{s}})$ and
$h_{y,i}^{(\sigma)}/(2K_{\mathrm{s}})$, the barrier takes the form
\begin{equation}
\Delta E_i
=
K_{\mathrm{s}} C_i \mu_i^{3/2},
\label{eq:barrier_generic}
\end{equation}
where $C_i$ is a dimensionless prefactor that depends on the location and
direction of approach to the spinodal boundary.

The longitudinal axis terminates at a cusp of the astroid, where the generic
fold expansion does not apply. For
$h_{y,i}^{(\sigma)}=0$ and $h_{x,i}>0$, define the axial detuning
\begin{equation}
\varepsilon_i
=
1-\frac{h_{x,i}}{2K_{\mathrm{s}}}.
\label{eq:axial_detuning}
\end{equation}
The spinodal margin and the axial detuning are related by
\begin{equation}
\mu_i
=
1-(1-\varepsilon_i)^{2/3},
\qquad
\varepsilon_i
=
1-(1-\mu_i)^{3/2}
=
\frac{3}{2}\mu_i+O(\mu_i^2).
\label{eq:margin_axial_relation}
\end{equation}
Expanding the conditional energy about the metastable binary state
$\theta_i=\pi$, with $q=\theta_i-\pi$, gives
\begin{equation}
E_i^{(\sigma)}(q)-E_i^{(\sigma)}(0)
=
K_{\mathrm{s}}\varepsilon_i q^2
-
\frac{K_{\mathrm{s}}}{4}q^4
+
O\!\left(\varepsilon_i q^4,q^6\right).
\label{eq:axis_cusp_expansion}
\end{equation}
The adjacent local maximum occurs at
\begin{equation}
q_{\mathrm{b}}^2
=
2\varepsilon_i+O(\varepsilon_i^2),
\end{equation}
and the corresponding barrier is
\begin{equation}
\Delta E_i
=
K_{\mathrm{s}}\varepsilon_i^2
+
O(K_{\mathrm{s}}\varepsilon_i^3)
=
\frac{9}{4}K_{\mathrm{s}}\mu_i^2
+
O(K_{\mathrm{s}}\mu_i^3).
\label{eq:barrier_axis}
\end{equation}

The barrier therefore vanishes as $\mu_i^{3/2}$ when the astroid is crossed
through a smooth nonaxial segment, but as $\mu_i^2$ when the longitudinal cusp
is approached.

\section{Stoner--Wohlfarth correspondence}
\label{sec:sw_correspondence}

The conditional energy derived in
Sec.~\ref{sec:spinodal_freezing} admits a direct correspondence with the
Stoner--Wohlfarth model of a single-domain magnetic particle with uniaxial
anisotropy~\cite{stoner1948mechanism,tannous2008stoner,braun2004exact}. Denoting the
anisotropy energy density by $K_u$, the particle volume by $V$, and the
saturation magnetization by $M_s$, the Stoner--Wohlfarth energy is
\begin{equation}
E_{\mathrm{SW}}(\vartheta)
=
K_uV\sin^2\vartheta
-
VM_sH_x\cos\vartheta
-
VM_sH_y\sin\vartheta,
\label{eq:sw-energy}
\end{equation}
where $\vartheta$ is the angle between the magnetization and the easy axis,
and $H_x$ and $H_y$ are the applied-field components parallel and
perpendicular to that axis, respectively.

Using
\begin{equation}
\sin^2\vartheta
=
\frac{1-\cos(2\vartheta)}{2},
\end{equation}
Eq.~\eqref{eq:sw-energy} becomes
\begin{equation}
\begin{split}
E_{\mathrm{SW}}(\vartheta)
=
\frac{K_uV}{2}
&-
\frac{K_uV}{2}\cos(2\vartheta)\\\\
&-
VM_sH_x\cos\vartheta
-
VM_sH_y\sin\vartheta.
\label{eq:sw-energy-cosine}
\end{split}
\end{equation}
The first term is independent of $\vartheta$ and therefore does not affect the stationary points or the shape of the energy landscape. Neglecting this additive constant, Eq.~\eqref{eq:sw-energy-cosine} has the same
functional form as the conditional oscillator energy
\begin{equation}
E_i^{(\sigma)}(\theta_i)
=
-\frac{K_{\mathrm{s}}}{2}\cos(2\theta_i)
-h_{x,i}\cos\theta_i
-h_{y,i}^{(\sigma)}\sin\theta_i.
\label{eq:local_field_energy_sw}
\end{equation}
The corresponding equivalence is
\begin{equation}
\begin{split}
\vartheta \leftrightarrow \theta_i,
\qquad
K_uV \leftrightarrow K_{\mathrm{s}},
\qquad\\\\
VM_sH_x \leftrightarrow h_{x,i},
\qquad
VM_sH_y \leftrightarrow h_{y,i}^{(\sigma)}.
\label{eq:sw_correspondence}
\end{split}
\end{equation}
The normalized correspondence can be established by introducing the
Stoner--Wohlfarth switching field
\begin{equation}
H_c=\frac{2K_u}{M_s}.
\end{equation}
Dividing the angle-dependent part of Eq.~\eqref{eq:sw-energy-cosine} by
$2K_uV$ gives
\begin{equation}
\widetilde{e}_{\mathrm{SW}}(\vartheta)
=
-\frac{1}{4}\cos(2\vartheta)
-
\frac{H_x}{H_c}\cos\vartheta
-
\frac{H_y}{H_c}\sin\vartheta.
\label{eq:sw-reduced}
\end{equation}
Likewise, normalizing Eq.~\eqref{eq:local_field_energy_sw} by
$2K_{\mathrm{s}}$ yields
\begin{equation}
\frac{E_i^{(\sigma)}(\theta_i)}{2K_{\mathrm{s}}}
=
-\frac{1}{4}\cos(2\theta_i)
-
\frac{h_{x,i}}{2K_{\mathrm{s}}}\cos\theta_i
-
\frac{h_{y,i}^{(\sigma)}}{2K_{\mathrm{s}}}\sin\theta_i.
\label{eq:oscillator-reduced}
\end{equation}
Thus,
\begin{equation}
\frac{H_x}{H_c}
\leftrightarrow
\frac{h_{x,i}}{2K_{\mathrm{s}}},
\qquad
\frac{H_y}{H_c}
\leftrightarrow
\frac{h_{y,i}^{(\sigma)}}{2K_{\mathrm{s}}}.
\label{eq:normalized_sw_correspondence}
\end{equation}
Under this correspondence, the SHI contribution maps onto the uniaxial anisotropy energy, while the instantaneous local network field maps onto the applied magnetic field. The spinodal condition derived for the oscillator landscape therefore recovers the Stoner--Wohlfarth switching astroid in the first-harmonic limit. 

\section{Conclusion}

We have developed a local energy-landscape description of bistability in phase-based Ising machines. The longitudinal component of the instantaneous network field sets the energy difference between the two binary phase states, whereas the transverse component leaves this splitting unchanged but shifts the stationary points and modifies the onset of bistability. Their competition with SHI produces an astroid spinodal boundary that determines the corresponding SHI threshold, including its dependence on field orientation. Near this boundary, the barrier height follows the generic fold scaling $\Delta E_i\propto\mu_i^{3/2}$, with the cusp scaling $\Delta E_i\propto\mu_i^{2}$ along the longitudinal axis. OIMs and DIMs share the same spinodal geometry, differing only by reflection of the transverse field. Finally, the first-harmonic conditional landscape is exactly equivalent, up to an additive constant, to the Stoner--Wohlfarth energy, placing local bistability in these machines within a well-established spinodal framework.


\section*{DATA AVAILABILITY} The data that support the findings of this study are available from the corresponding author upon reasonable request.
\section*{Acknowledgment}
The authors thank  Hasantha Ekanayake and Nikhat Khan for their valuable inputs. This material is based upon work supported in part by National Science Foundation grant \#No. 2328961 .
\clearpage

\clearpage

\bibliography{References}

@article{mohseni2022ising,
  author  = {Mohseni, Naeimeh and McMahon, Peter L. and Byrnes, Tim},
  title   = {Ising Machines as Hardware Solvers of Combinatorial Optimization Problems},
  journal = {Nature Reviews Physics},
  volume  = {4},
  pages   = {363--379},
  year    = {2022},
  url     = {https://doi.org/10.1038/s42254-022-00440-8}
}

@article{zhang2024review,
  author  = {Zhang, Tingting and Tao, Qichao and Liu, Bailiang and Grimaldi, Andrea and Raimondo, Eleonora and Jim{\'e}nez, Manuel and Avedillo, Mar{\'\i}a Jos{\'e} and Nunez, Juan and Linares-Barranco, Bernab{\'e} and Serrano-Gotarredona, Teresa and Finocchio, Giovanni and Han, Jie},
  title   = {A Review of {Ising} Machines Implemented in Conventional and Emerging Technologies},
  journal = {IEEE Transactions on Nanotechnology},
  volume  = {23},
  pages   = {704--717},
  year    = {2024},
  url     = {https://doi.org/10.1109/TNANO.2024.3457533}
}

@article{lucas2014ising,
  author    = {Lucas, Andrew},
  title     = {Ising Formulations of Many {NP} Problems},
  journal   = {Frontiers in Physics},
  volume    = {2},
  pages     = {74887},
  year      = {2014},
  publisher = {Frontiers},
  url       = {https://doi.org/10.3389/fphy.2014.00005}
}

@article{chang2024quantum,
  author  = {Chang, Yen-Jui and Nien, Chin-Fu and Huang, Kuei-Po and Zhang, Yun-Ting and Cho, Chien-Hung and Chang, Ching-Ray},
  title   = {Quantum Computing for Optimization with {Ising} Machine},
  journal = {IEEE Nanotechnology Magazine},
  volume  = {18},
  number  = {3},
  pages   = {15--22},
  year    = {2024},
  url     = {https://doi.org/10.1109/MNANO.2024.3378485}
}

@article{mcmahon2016fully,
  author    = {McMahon, Peter L. and Marandi, Alireza and Haribara, Yoshitaka and Hamerly, Ryan and Langrock, Carsten and Tamate, Shuhei and Inagaki, Takahiro and Takesue, Hiroki and Utsunomiya, Shoko and Aihara, Kazuyuki and others},
  title     = {A Fully Programmable 100-Spin Coherent {Ising} Machine with All-to-All Connections},
  journal   = {Science},
  volume    = {354},
  number    = {6312},
  pages     = {614--617},
  year      = {2016},
  publisher = {American Association for the Advancement of Science},
  url       = {https://doi.org/10.1126/science.aah5178}
}

@article{ezawa2022ising,
  author  = {Ezawa, Motohiko and Lebrasseur, Eric and Mita, Yoshio},
  title   = {{Ising} Machine Based on Bistable Microelectromechanical Systems},
  journal = {Journal of the Physical Society of Japan},
  volume  = {91},
  number  = {11},
  pages   = {114601},
  year    = {2022},
  url     = {https://doi.org/10.7566/JPSJ.91.114601}
}

@article{grimaldi2023evaluating,
  author  = {Grimaldi, Andrea and Mazza, Luciano and Raimondo, Eleonora and Tullo, Pietro and Rodrigues, Davi and Camsari, Kerem Y. and Crupi, Vincenza and Carpentieri, Mario and Puliafito, Vito and Finocchio, Giovanni},
  title   = {Evaluating Spintronics-Compatible Implementations of {Ising} Machines},
  journal = {Physical Review Applied},
  volume  = {20},
  number  = {2},
  pages   = {024005},
  year    = {2023},
  url     = {https://doi.org/10.1103/PhysRevApplied.20.024005}
}

@article{chou2019analog,
  author    = {Chou, Jeffrey and Bramhavar, Suraj and Ghosh, Siddhartha and Herzog, William},
  title     = {Analog Coupled Oscillator Based Weighted {Ising} Machine},
  journal   = {Scientific Reports},
  volume    = {9},
  number    = {1},
  pages     = {14786},
  year      = {2019},
  publisher = {Nature Publishing Group UK London},
  url       = {https://doi.org/10.1038/s41598-019-49699-5}
}

@article{mallick2023cmos,
  author  = {Mallick, Antik and Zhao, Zijian and Bashar, Mohammad Khairul and Alam, Shamiul and Islam, Md Mazharul and Xiao, Yi and Xu, Yixin and Aziz, Ahmedullah and Narayanan, Vijaykrishnan and Ni, Kai and others},
  title   = {{CMOS}-Compatible {Ising} Machines Built Using Bistable Latches Coupled through Ferroelectric Transistor Arrays},
  journal = {Scientific Reports},
  volume  = {13},
  pages   = {1515},
  year    = {2023},
  url     = {https://doi.org/10.1038/s41598-023-28192-0}
}

@article{bashar2021experimental,
  author  = {Bashar, Mohammad Khairul and Mallick, Antik and Shukla, Nikhil},
  title   = {Experimental Investigation of the Dynamics of Coupled Oscillators as {Ising} Machines},
  journal = {IEEE Access},
  volume  = {9},
  pages   = {148184--148190},
  year    = {2021},
  url     = {https://doi.org/10.1109/ACCESS.2021.3124808}
}

@article{moy20221,
  author    = {Moy, William and Ahmed, Ibrahim and Chiu, Po-wei and Moy, John and Sapatnekar, Sachin S. and Kim, Chris H.},
  title     = {A 1,968-Node Coupled Ring Oscillator Circuit for Combinatorial Optimization Problem Solving},
  journal   = {Nature Electronics},
  volume    = {5},
  number    = {5},
  pages     = {310--317},
  year      = {2022},
  publisher = {Nature Publishing Group UK London},
  url       = {https://doi.org/10.1038/s41928-022-00749-3}
}

@inproceedings{afoakwa2021brim,
  author       = {Afoakwa, Richard and Zhang, Yiqiao and Vengalam, Uday Kumar Reddy and Ignjatovic, Zeljko and Huang, Michael},
  title        = {{BRIM}: Bistable Resistively-Coupled {Ising} Machine},
  booktitle    = {2021 IEEE International Symposium on High-Performance Computer Architecture (HPCA)},
  pages        = {749--760},
  year         = {2021},
  organization = {IEEE},
  url          = {https://doi.org/10.1109/HPCA51647.2021.00068}
}

@article{Cassella,
  title = {Parametric Frequency Divider Based Ising Machines},
  author = {Casilli, Nicolas and Kaisar, Tahmid and Colombo, Luca and Ghosh, Siddhartha and Feng, Philip X.-L. and Cassella, Cristian},
  journal = {Phys. Rev. Lett.},
  volume = {132},
  issue = {14},
  pages = {147301},
  numpages = {7},
  year = {2024},
  month = {Apr},
  publisher = {American Physical Society},
  doi = {10.1103/PhysRevLett.132.147301},
  url = {https://link.aps.org/doi/10.1103/PhysRevLett.132.147301}
}

@inproceedings{wang2019oim,
  author    = {Wang, Tianshi and Roychowdhury, Jaijeet},
  title     = {{OIM}: Oscillator-Based {Ising} Machines for Solving Combinatorial Optimisation Problems},
  booktitle = {International Conference on Unconventional Computation and Natural Computation},
  pages     = {232--256},
  year      = {2019},
  publisher = {Springer},
  url       = {https://doi.org/10.1007/978-3-030-19311-9_19}
}

@article{ekanayake2025different,
  author  = {Ekanayake, E. M. H. E. B. and Shukla, Nikhil},
  title   = {Different Paths, Same Destination: Designing Physics-Inspired Dynamical Systems with Engineered Stability to Minimize the {Ising} Hamiltonian},
  journal = {Physical Review Applied},
  volume  = {24},
  pages   = {024008},
  year    = {2025},
  url     = {https://doi.org/10.1103/cdc9-y234}
}

@article{restrepo2005onset,
  author  = {Restrepo, Juan G. and Ott, Edward and Hunt, Brian R.},
  title   = {Onset of Synchronization in Large Networks of Coupled Oscillators},
  journal = {Physical Review E},
  volume  = {71},
  number  = {3},
  pages   = {036151},
  year    = {2005},
  url     = {https://doi.org/10.1103/PhysRevE.71.036151}
}

@article{stoner1948mechanism,
  author    = {Stoner, Edmund Clifton and Wohlfarth, Ernest P.},
  title     = {A Mechanism of Magnetic Hysteresis in Heterogeneous Alloys},
  journal   = {Philosophical Transactions of the Royal Society of London. Series A, Mathematical and Physical Sciences},
  volume    = {240},
  number    = {826},
  pages     = {599--642},
  year      = {1948},
  publisher = {The Royal Society London},
  url       = {https://doi.org/10.1098/rsta.1948.0007}
}

@article{tannous2008stoner,
  author  = {Tannous, C. and Gieraltowski, J.},
  title   = {The {Stoner}--{Wohlfarth} Model of Ferromagnetism},
  journal = {European Journal of Physics},
  volume  = {29},
  number  = {3},
  pages   = {475--487},
  year    = {2008},
  url     = {https://doi.org/10.1088/0143-0807/29/3/008}
}

@article{braun2004exact,
  author    = {Braun, Daniel},
  title     = {Exact Activation Energy of Magnetic Single Domain Particles},
  journal   = {Journal of Magnetism and Magnetic Materials},
  volume    = {283},
  number    = {1},
  pages     = {1--7},
  year      = {2004},
  publisher = {Elsevier},
  url       = {https://doi.org/10.1016/j.jmmm.2004.05.007}
}

\end{document}